\documentclass[12pt,a4paper]{article}

\usepackage{graphicx,here}

\def\Be{\begin{equation}}
\def\Ee{\end{equation}}
\def\KeffS{K_{\mbox{\footnotesize eff}}^{\mbox{\footnotesize[S]}}}
\def\KeffW{K_{\mbox{\footnotesize eff}}^{\mbox{\footnotesize[W]}}}
\def\etaC{\eta_{\mbox{\footnotesize c}}}

\begin{document}

\begin{center}

{\Large Finite-Range Scaling Method to Analyze Systems\\
with Infinite-Range Interactions
}

\vspace{5mm}

{\small Ken-Ichi Aoki \footnote{aoki@hep.s.kanazawa-u.ac.jp}, Tamao Kobayashi \footnote{ballblue@hep.s.kanazawa-u.ac.jp}
and Hiroshi Tomita \footnote{t\_hirosi@hep.s.kanazawa-u.ac.jp}}

\vspace{5mm}

{\small \it Institute for Theoretical Physics, Kanazawa University,\\
Kamuma, Kanazawa, 920-1192, Japan}

\vspace{5mm}

{\bf \small abstract}
\end{center}

{\small We propose a new practical method for evaluating the critical coupling constant in 
one-dimensional long-range interacting systems. 
We assume a finite-range scaling and define its exponent 
for the logarithm of the susceptibility. We 
find criticality in the form of a zeta function singularity.
As an example, we present results for a long-range Ising model.}

\vspace{10mm}

Dissipative quantum mechanics has drawn attention for a long time.
Recently, experiments observing quantum decoherence have enhanced the interest in such systems.
We need a deeper theoretical understanding to describe decoherence.

We start with a dynamical system in which the target variable is 
surrounded by many environmental degrees of
freedom.\cite{Caldeira-Leggett} Then, energy flow
from the target system to the environment 
might appear as energy dissipation effects.
To model such behavior, we consider a set of infinitely many harmonic oscillators linearly coupled to the
target system and integrate over these environmental degrees of freedom. 
We thereby obtain effective interactions for the target system variable, which are, in general, of infinitely long range and may effectively work as dissipation.

Decoherence physics has been studied with a double-well potential by 
observing Rabi oscillation, which is driven by quantum coherence.
Due to dissipative effects, decoherence appears to suppress the oscillation.
This is seen as tunneling suppression, or a localization 
phase transition, due to long-range interactions.
Such systems have been studied by many authors, typically using, 
canonical methods,\cite{Fujikawa-Iso-Sasaki-Suzuki92}
the non-perturbative renormalization group,\cite{NPRG}
and sophisticated Monte Carlo simulations.\cite{MC-DW06} 
It is very difficult to fully analyze systems of the type in question and
they are often approximated by the smallest number of degrees of freedom, 
that is, a two-state approximation. 
In this case, the system is merely a one-dimensional Ising model.

The long-range Ising model has its own long history of research. 
Here, we only cite the original works\cite{Ruelle68,Dyson69} 
and one recent work.\cite{Aizenman-Fernandez88}
If the interaction range is finite, 
then the system has no phase transition.
However, when the interaction range is infinite and sufficiently strong, there can be a phase
transition giving rise to spontaneous magnetization.
Actually, it is not easy to evaluate the critical coupling constant, and this requires a
large simulation, even in the Ising case.\cite{MC-Ising01}

The aim of this paper is to briefly present a new, practical method for evaluating the 
critical coupling constant in the case of interactions with infinitely long range. 
First, we limit the range of interactions to some finite value $n$, and then we solve the system precisely by using
an extended type of decimation renormalization group method which we call the
block decimation renormalization group (BDRG).
Then, we assume a scaling relation, referred to as finite range scaling (FRS).
We define an exponent characterizing the range $n$ dependence of physical quantities and evaluate the exponent. 
Using the obtained FRS exponent, we estimate the properties of the system in the case of an infinite range. In this treatment, there appears the zeta function that determines the criticality.
This FRS method is similar to the finite-size scaling method used in simulations of finite lattice systems to estimate the properties of infinitely large systems.
The FRS method can be applied to any system with infinite-range interactions in the case that the system with finite-range cuttoff can be solved effectively.
Application of this method to double-well dissipative quantum mechanics will be reported
elsewhere.\cite{Aoki-Kobayashi-Tomita08a}

We consider the one-dimensional Ising model, whose action is
defined by
\Be
S = - \sum_{i;j>1}K_j\sigma_i\sigma_{i+j} - h \sum_i\sigma_i\ ,
\Ee
where each spin variable, $\sigma_i$, takes the value $1$ or $-1$, and $h$ is an external field.
The coupling constants, $K_j$, assumed to be non-negative, determine the interactions
between spins separated by a distance $j$.

As a first step, we study the nearest-neighbor model, in which only $K_1$ is non-vanishing.
The interactions can be represented by a $2\times 2$ matrix
\Be
T^{(0)}= 
\left[ 
\begin{array}{cc}
\exp(K_1+h) & \exp(-K_1)\\
\exp(-K_1) &  \exp(K_1-h)\\
\end{array} 
\right]\ .
\Ee
We define the decimation renormalization group (DRG) transformation by integrating out all the 
even site spins and reducing effective interactions among the odd site spins.
These effective interactions also take a nearest-neighbor form.
Thus this renormalization procedure can be iterated, and we define the $k$-th renormalized
interactions $T^{(k)}$ by induction,
\Be
T^{(k)} \equiv T^{(k-1)}\ T^{(k-1)}\ ,
\Ee
which are interactions between spins on sites separated by a distance of $2^k$.
Then, we define the susceptibility with respect to $h$ for the infinite system as
\Be
\chi \equiv \lim_{k\rightarrow \infty}
\left. \frac{1}{2^k}\frac{\partial^2}{\partial h^2} \log \mbox{Tr}\  T^{(k)} \right|_{h=0}\ .
\Ee
In this simplest case, the susceptibility is exactly obtained as $\exp(2K_1)$.

Next, we include non-nearest-neighbor interactions, but we still 
limit the range of the interaction to $n$, that is, 
$K_j =0$ for $j>n$. 
This is the first stage in the FRS method.
The standard DRG is not effective in such a situation. 
To carry out the computation for this system, we divide spins into blocks of size $n$.
Then, the interactions are all limited to ``nearest-neighbor" inter-block interactions.
Thus, the system is a one-dimensional nearest neighbor block-spin system.

One block contains $n$ spins, and it possesses $2^n$ states.
The inter-block interactions are represented by a $2^n \times 2^n$ matrix.
A renormalization group transformation, 
which we call the block decimation renormalization group (BDRG)\cite{Aoki-Kobayashi-Tomita08a},  is defined for this block-spin matrix.
Numerical calculation of the BDRG gives a precise value of the susceptibility for the
system with interactions of range $n$, which we denote $\chi(n)$.

Here, we propose two inequalities for the logarithm of the susceptibility,
\Be
2\KeffW \le \log \chi \le 2\KeffS\ .
\Ee
Two effective coupling constants, that are appropriate for the weak and strong regions, respectively, are defined by
\Be
\KeffW \equiv \sum_j K_j \ ,\ \ \KeffS \equiv \sum_j j K_j\ . \label{eq:KeffWS_def}
\Ee
They have been called the 0th and 1st moments of the coupling constants in references and have played crucial roles in determining phase structures.
We conjecture that these inequalities hold for any set of non-negative coupling constants $K_j$.
Though we have not yet succeeded in strictly proving these inequalities, we have found no exceptions
 in our calculation employing BDRG.

Physically, we can understand these bounds as follows: 
The weak bound, $2\KeffW$, comes from the
1st-order perturbation expansion of $\log \chi$, while the strong bound 
comes from the case of an almost ordered system, for which an approximate BDRG equation provides an accurate description and  gives 
the boundary value $2\KeffS$.\cite{Aoki-Kobayashi-Tomita08a}
In case of nearest-neighbor interactions, both equalities hold exactly.

Hereafter, we consider systems whose 
coupling constants $K_j$ decrease with $j$ according to a power low, in the form
\Be
K_j = \frac{\eta}{j^p}\ . \label{eq:K_ohmic}
\Ee
Such power damping behavior is expected to appear in the case of quantum 
dissipation due to the influence of the environmental degrees of freedom. In fact, if we set $p=2$, the 
resultant effective dissipation is equivalent to that in the case that there is a friction force proportional
to the velocity.\cite{Caldeira-Leggett}
For this reason, this case is called the Ohmic case.
For the above coupling constants in Eq.~(\ref{eq:K_ohmic}), the inequalities read
\Be
2\eta\zeta(p) \le \log\chi \le 2\eta\zeta(p-1)\ ,
\label{eq:ineq-zeta-p}
\Ee
where $\zeta$ is the standard zeta function.
For example, the result of the BDRG calculation of the susceptibility for $n=11, p=1.8$ and $\eta=[0,1]$ is
displayed in Fig.~1 with the lower and upper bounds defined in Eq.~(\ref{eq:KeffWS_def}).
That figure clearly shows that the logarithm of the susceptibility is accurately described by these 
boundary lines in the weak and strong coupling regions, respectively, and it moves 
from the lower bound to the upper bound rather quickly.

\begin{figure}[htb]
\parbox{0.49\textwidth}{
\includegraphics[width=6.6cm]{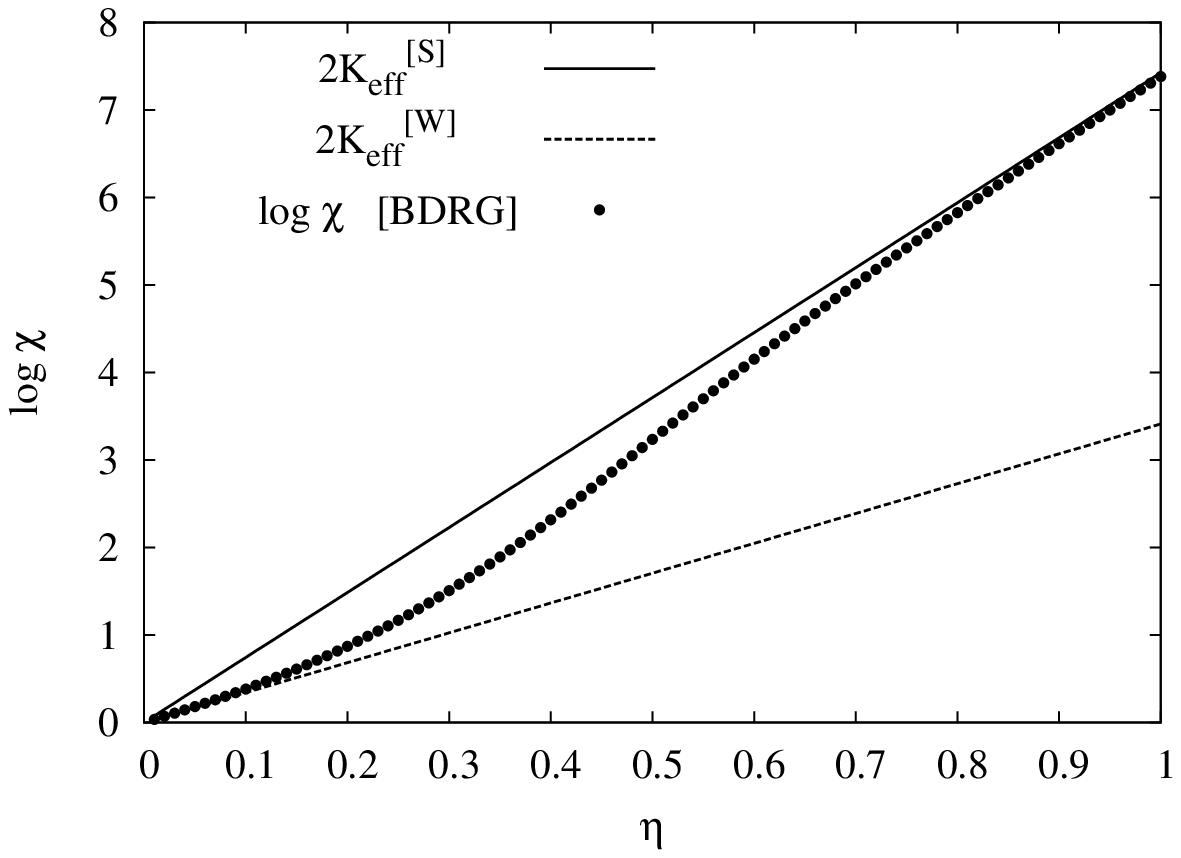}
\caption{The susceptibility for $p=1.8,\eta=[0,1]$ and $n=11$, with the upper and lower bounds.}
\label{fig:chi-eta-p1.8-n11-bound}}
\hfill
\parbox{0.49\textwidth}{
\includegraphics[width=6.6cm]{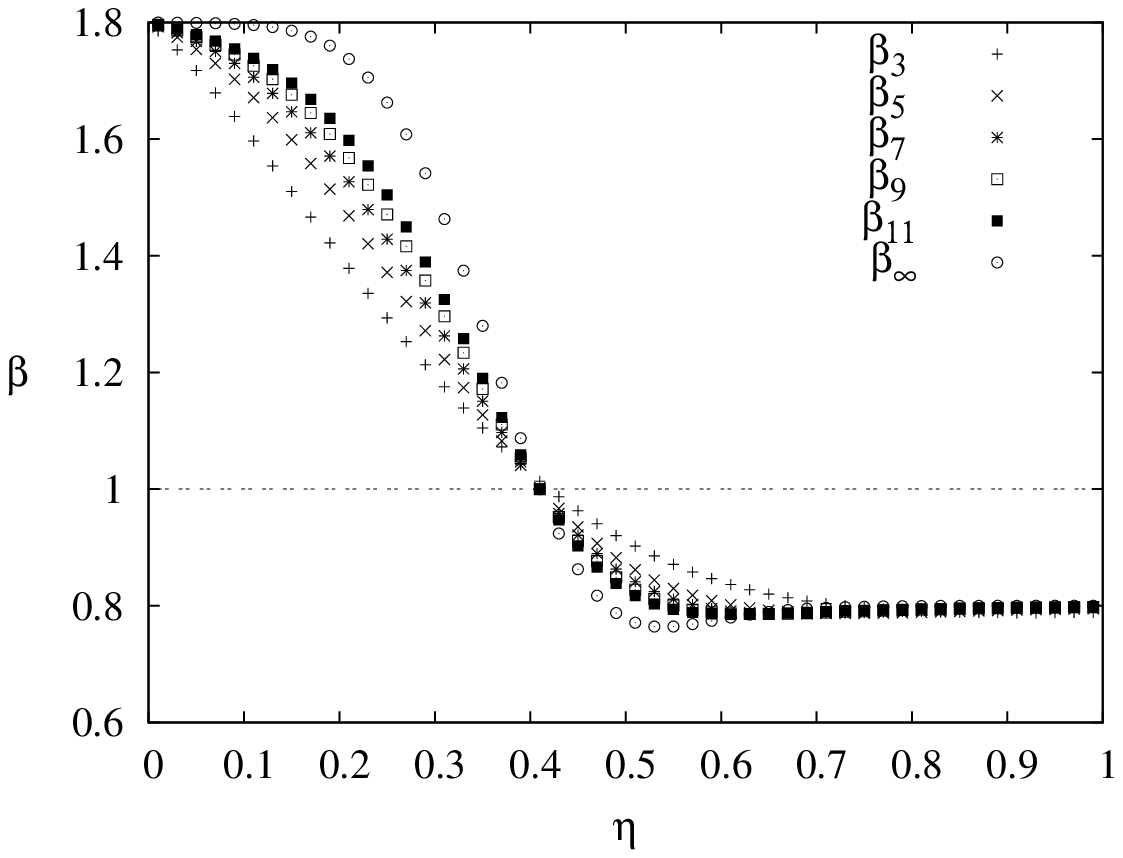}
\caption{The FRS exponent $\beta$ for $p=1.8,\hfil\break \eta=[0,1]$ and $n=3,5,7,9,11,\infty$.}
\label{fig:chi-eta-p1.8-ext11}}
\end{figure}

Now, we describe the FRS method.
We calculate the change in the logarithm of the susceptibility when the range of the interaction changes by 1:
\Be
\Delta(n,p,\eta)\equiv\frac{1}{2\eta}\left[\log\chi(n) - \log\chi(n-1)\right]\ .
\Ee
At the boundaries, this should behave as 
\Be
\Delta^{\mbox{\footnotesize[W]}} = \left(\frac{1}{n}\right)^{p}\ ,\ \ 
\Delta^{\mbox{\footnotesize[S]}} = \left(\frac{1}{n}\right)^{p-1}\ .
\Ee
Considering these boundary forms, we define the FRS
exponent for $\Delta$ by
\Be
\beta(p,\eta)\equiv\lim_{n\rightarrow \infty}
\frac{\log\Delta(n,p,\eta)}{-\log n} \ .
\Ee
We have not proved the existence of this limit. It is plausible, however, that it does exist.
Then, the infinite-range contribution to the logarithm of the susceptibility 
is represented by the zeta function,
\Be
\lim_{n\rightarrow\infty}\log \chi (n) \propto \zeta(\beta(p,\eta))\ .
\Ee
As a real function, the zeta function $\zeta(z)$ is finite
for $z>1$ and diverges as $z\rightarrow 1_+$. 
Thus, $\beta(p,\eta)=1$ corresponds to
 criticality.

On the other hand, from the inequalities (\ref{eq:ineq-zeta-p}), 
we have the inequalities for the FRS exponent
\Be
p \ge \beta(p,\eta) \ge p-1\ .
\Ee
Because $\chi (n)$ is monotonic with respect to $n$\cite{Griffiths67a}, $\beta(p,\eta)$ is monotonic with respect to $\eta$, and $\beta(p,\eta)$ changes from $p$ to $p-1$ as $\eta$ changes from 0 to $\infty$.

Taking account of these properties of the FRS exponent, 
we can draw some conclusions about the properties of the phase transition.
For $p<1$, the susceptibility cannot be finite as $n\rightarrow \infty$ for any 
$\eta\ne0$. Thus, the system is always ordered for any $\eta\ne0$.
By contrast, for $p>2$, the susceptibility is always finite 
as $n\rightarrow \infty$ for any finite $\eta$. Thus, in this case, the ordered phase never appears.
For the intermediate region satisfying $1<p\le 2$, there can be a phase transition 
point at which $\beta(p,\eta)$ crosses the value 1.
These basic properties have been known for many years, and they were originally demonstrated by other reasoning or more rigorous arguments.\cite{Ruelle68,Dyson69} 
Our aim here is to evaluate the critical coupling constant
$\etaC$ ($p$).

Now we report the results of numerical calculations of the FRS exponent to determine $\eta_{\mbox{\footnotesize c}}$.
In the actual evaluation of the exponent, we employed the formula
\Be
\beta(n,p,\eta) \equiv - \frac{\log \Delta(n,p,\eta) - \log \Delta(n-1,p,\eta)}
                         {\log n - \log (n-1)}
\Ee
to define $\beta(n)$.
Analyzing its $n$ dependence, we estimate the value of $\beta$ in the $n \rightarrow \infty$ limit by 
linearly extrapolating with respect to $1/n$.
With the computing resource of a desktop computer, the size $n=11$ is the practical limit for 
an overnight calculation, and hence, in this paper, we consider data up to $n=11$.

In Fig.~2, we plot an example of the $n$ dependence analysis for 
$\beta(n, p=1.8,\eta)$, with $n=3,5,7,9,11,\infty$ (which is the extrapolated
value from $n=10$ and $11$).
According to the FRS exponent inequalities, $\beta$ changes from  the weak limit, 1.8, to the strong limit, 0.8. 
In the region of this monotonic decrease, there exists the point $\beta=1$, corresponding to the critical coupling constant $\etaC$, which 
is found to be about 0.41.
From the data, the decrease is not strictly monotonic, and the lower bound is broken, especially when we consider linearly extrapolated values. 
However, we believe that this is spurious behavior, due to the fact that $n$ is too small.

We should also note that around the most important region, $\beta\simeq 1$, $\beta(n)$ is 
nearly independent of $n$, which is rather surprising, since the critical
point $\etaC$ is correctly found even with $n=3$.
This stability near the criticality is seen for all values of $p$.
Contrastingly, linear extrapolation is necessary 
for the weak and strong
regions, and this makes the transition sharper.
We conjecture that at the critical point, the function $\beta(\eta)$ is singular 
with an infinite derivative and may in fact jump to the strong limit value.

Calculating the FRS exponent $\beta$ for $n=9, p=[0.9,2.1], \eta=[0,1]$, 
we obtain the contour plot given in Fig.~3. 
We plot $\beta(n=9)$ without extrapolation. 
The thick curve represents the contour of $\beta=1$, and the adjacent curves represent the contours
with spacings of 0.2. 
The $\beta=1$ curve is the phase boundary in the 
$p$-$\eta$ plane. The lower side is the symmetric phase, and the upper side is the
symmetry broken phase. We see  that the boundary values of the $\beta$ inequalities
accurately approximate the weak and strong region behavior for all $p$.
We also see that non-monotonic behavior is stronger for smaller $p$, but it
does not seem to affect the critical region.

\begin{figure}[htb]
\parbox{0.49\textwidth}{
\includegraphics[width=6.8cm]{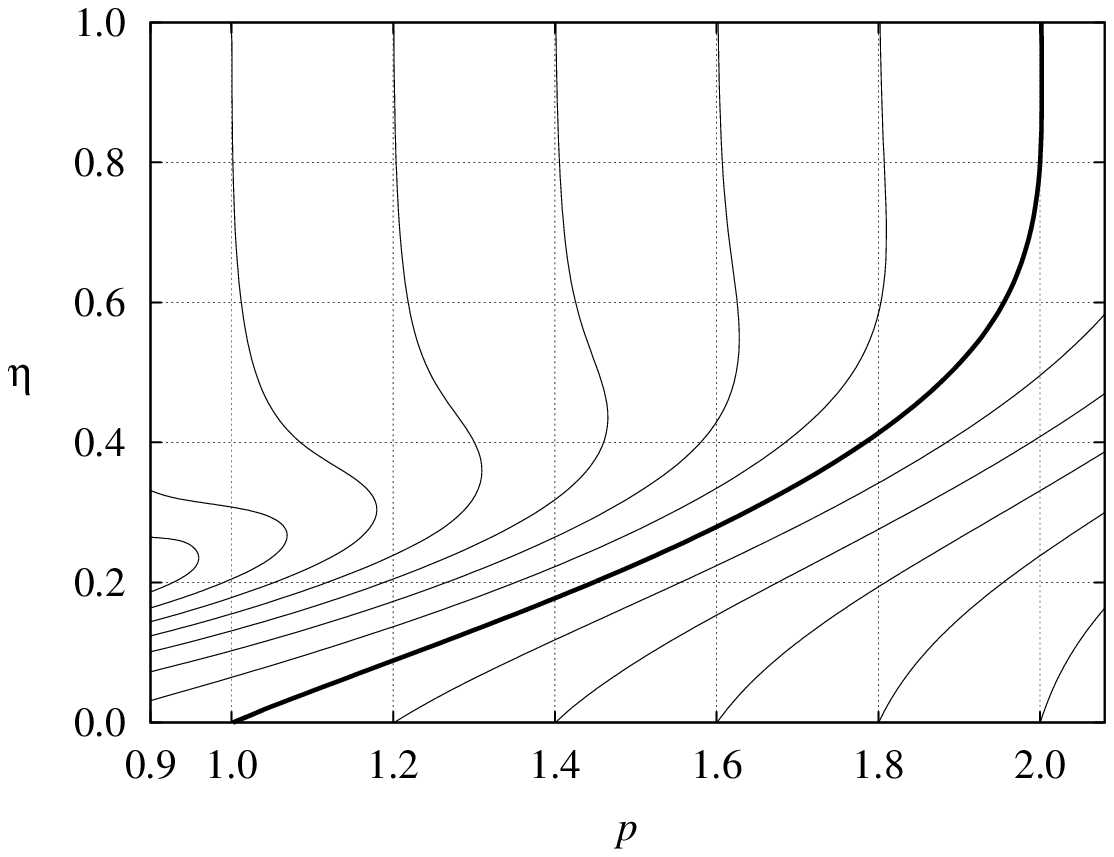}
\caption{Contour map of the FRS exponent $\beta$ for $n=9, p=[0.9,2.1], \eta=[0,1]$.}
\label{fig:beta-p-eta-contour-line}
}\hfill
\parbox{0.49\textwidth}{
\includegraphics[width=6.8cm]{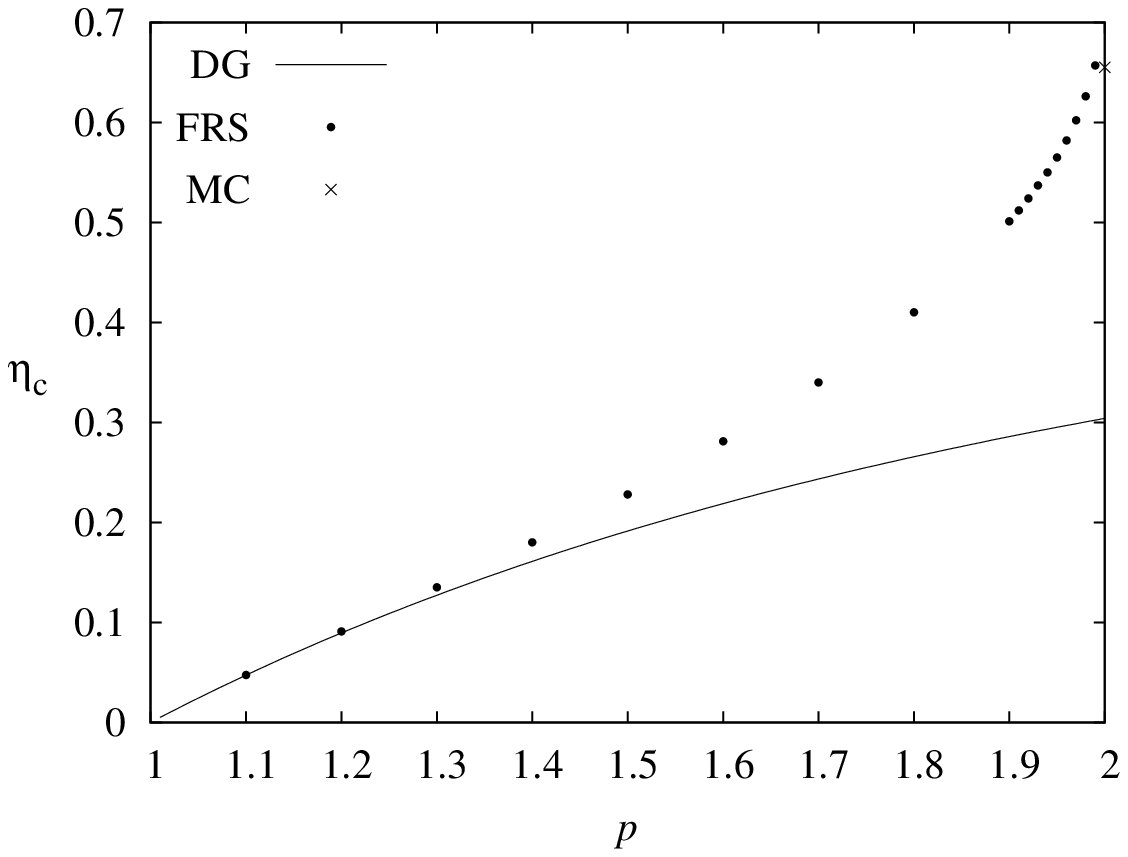}
\caption{The critical coupling constant $\etaC$ as a function of $p$.}
\label{fig:etac-p-comparison}}
\end{figure}

\begin{table}[htb]
\begin{center}
\caption{The critical coupling constant $\etaC(p)$}
\begin{tabular}{|r|r||r|r||r|r||r|r|}
\hline
$p\hskip3mm$& $\etaC\hskip5mm$ & $p\hskip3mm$\  &$\etaC\hskip5mm$
&$p\hskip3mm$\  &$\etaC\hskip5mm$ &$p\hskip3mm$\  &$\etaC\hskip5mm$ \\
\hline\hline
1.10&0.0474(1)&1.60&0.281(1)&1.91&0.512(1)&1.96&0.582(1)\\\hline
1.20&0.091(1)& 1.70&0.340(1)&1.92&0.524(1)&1.97&0.602(1)\\\hline
1.30&0.135(1)& 1.80&0.410(1)&1.93&0.537(1)&1.98&0.626(1)\\\hline
1.40&0.180(1)& 1.90&0.501(1)&1.94&0.550(1)&1.99&0.657(2)\\\hline
1.50&0.228(1)&     &        &1.95&0.565(1)&    &        \\\hline
\end{tabular}
\end{center}
\end{table}

Values for the critical coupling constant, $\eta_c(p)$, are listed in
Table I. For $p=2$, it is not easy to evaluate $\etaC$ with high precision, 
and therefore here we list results for $p\le 1.99$.
Figure 4 compares our results with the strict lower bound obtained by
Dyson and Griffiths\cite{Dyson69,Griffiths67}(DG), $\etaC(p) > 1/(2\zeta(p))$, 
and with recent high precision Monte Carlo simulation results (MC)
for $p=2$, $\etaC=0.6551(6)$.\cite{MC-Ising01} 
It is seen that our results do not go beyond the DG bound, and actually, they are very near
to each other in the region of small $p$. The MC results and our results appear to be consistent, but the region near $p=2$ should be examined in more detail.

In this paper, we do not compute the critical exponents of the phase transition.
The $\zeta(z)$ function has a simple pole at $z=1$, and we can deduce the
susceptibility critical exponent and its transition type (standard or
Kosterlitz-Thouless) by examining the details of the $\beta(p,\eta)$  behavior near $\etaC$. 

We would like to thank  Atsushi Horikoshi for fruitful discussions and Takashi Hara and Keiichi~R.~Ito for valuable suggestions.
This research was partially supported by the Ministry of Education, 
Science, Sports and Culture through a Grant-in-Aid for 
Scientific Research (B)(No.17340070, 2007).


\begin{thebibliography}{99}

\bibitem{Caldeira-Leggett}
A.~O.~Caldeira and A.~J.~Leggett, Phys. Rev. Lett. {\bf 46} (1981), 211; Ann. of Phys. {\bf 149} (1983), 374.

\bibitem{Fujikawa-Iso-Sasaki-Suzuki92}
K.~Fujikawa, S.~Iso, M.~Sasaki and H.~Suzuki, Phys. Rev. Lett. {\bf 68} (1992), 1093; Phys. Rev. B {\bf 46} (1992), 10295.

\bibitem{NPRG}
K-I.~Aoki, Int. J. Mod. Phys. {\bf 14} (2000), 1249; \\
K-I.~Aoki, A.~Horikoshi, M.~Taniguchi and H.~Terao, Phys. Rev. Lett. {\bf 108} (2002), 572;\\
K-I.~Aoki and A.~Horikoshi, Phys. Lett. A {\bf 314} (2003),177; Phys. Rev. A {\bf 66} (2002), 042105.


\bibitem{MC-DW06}
T.~Matsuo, Y.~Natsume and T.~Kato, J. Phys. Soc. Jpn.{\bf 75} (2006), 103002.

\bibitem{Ruelle68}
D.~Ruelle, Commun. Math. Phys.{\bf 9} (1968), 267.

\bibitem{Dyson69}
F.~J.~Dyson, Commun. Math. Phys. {\bf 12} (1969), 91.

\bibitem{Aizenman-Fernandez88}
M.~Aizenman and R.~Fern\'andez, Lett.~Math.~Phys.{\bf 16} (1988), 39.

\bibitem{MC-Ising01}
E.~Luijten and H.~Me\ss ingfeld, Phys. Rev. Lett. {\bf 86} (2001), 5305.

\bibitem{Aoki-Kobayashi-Tomita08a}
K-I.~Aoki, T.~Kobayashi and H.~Tomita, in preparation.

\bibitem{Griffiths67a}
R.~B.~Griffiths, J. Math. Phys. {\bf 8} (1967), 478.

\bibitem{Griffiths67}
R.~B.~Griffiths, Commun. Math. Phys.{\bf 6} (1967), 121.


\end{thebibliography}
\end{document}